# Charge Density Waves in Exfoliated Thin Films of *Van der Waals* Materials

Pradyumna Goli, Javed Khan, Darshana Wickramaratne, Roger K. Lake and Alexander A. Balandin[*]

*Department of Electrical Engineering and Materials Science and Engineering Program, Bourns College of Engineering, University of California – Riverside, Riverside, California 92521 USA*

## Abstract

A number of the charge-density-wave materials reveal a transition to the macroscopic quantum state around 200 K. We used graphene-like mechanical exfoliation of $TiSe_2$ crystals to prepare a set of films with different thicknesses. The transition temperature to the charge-density-wave state was determined via modification of Raman spectra of $TiSe_2$ films. It was established that the transition temperature can increase from its bulk value to ~240 K as the thickness of the van-der-Waals films reduces to the nanometer range. The obtained results are important for the proposed applications of such materials in the collective-state information processing, which require room-temperature operation.

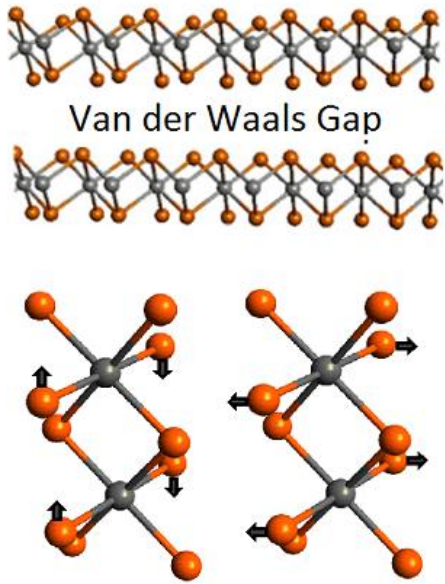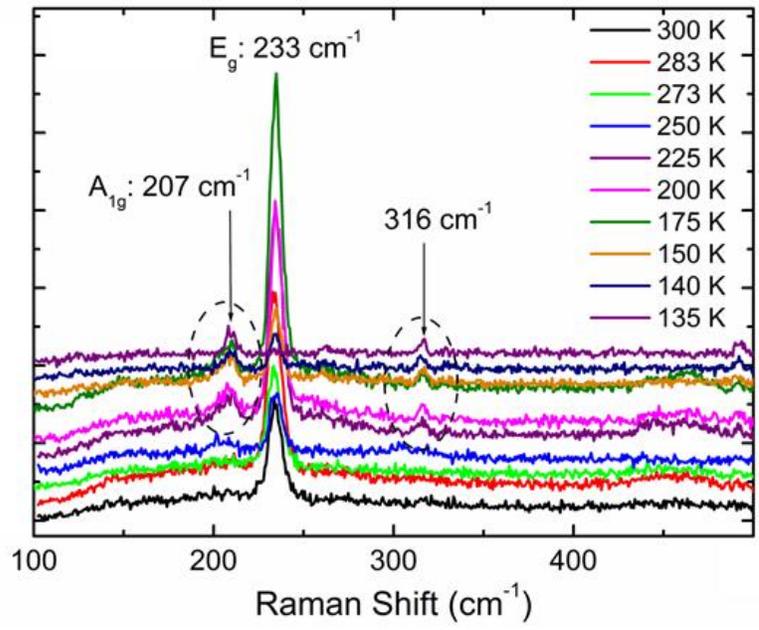

**Contents Image**

A charge density wave (CDW) is a symmetry-reducing ground state most commonly found in layered materials [1]. The appearance of a CDW state results from a Peierls instability [2]. Below the transition temperature, $T_C$, the lattice of atoms undergoes a periodic distortion and the electrons condense into a ground state with a periodic modulation of the charge density leading to an energy gap at the Fermi surface (Figure 1). For small applied electric fields, the CDW remains pinned to defects of the underlying lattice. Above a threshold field, $E_T$, the CDW can de-pin from the defects and slide through the crystal producing a collective current. CDW materials have been considered for possible use in electronic and optoelectronic devices [3-4]. The current modulation by the gate bias was studied in CDW field-effect transistors (FET). Field-effect modulation of CDW transport was demonstrated in $NbSe_3$ and $TaS_3$ FETs [5]. At the temperatures below the Peierls transition, an applied gate voltage modulated the collective conductance in CDW devices by more than two orders of magnitude larger than the single-particle conductance [5].

We have recently proposed the use of CDW collective states as alternative physical state variables for information processing [6]. It has become clear that power dissipation is the limiting factor to the continued scaling of size and speed of complementary metal-oxide semiconductor (CMOS) transistors [7]. The thermodynamic energy dissipation limit for one switching cycle of a charge-based logic bit is $Nk_BT\times ln(2)$ [8-9], where $N$ is the number of electrons, $k_B$ is Boltzmann's constant, and $T$ is the temperature. The assumption underlying this limit is that the electrons or spins act as an ensemble of *independent* particles. If instead, the $N$ electrons are in a *collective* state, then the minimum dissipation limit for one switching cycle can be reduced from $Nk_BT\times ln(2)$ to $k_BT\times ln(2)$ [9]. This fundamental fact provides a strong motivation to exploit collective states as alternative state variables for information processing. CDW materials reveal collective states with the macroscopic quantum coherence length. The collective states that exist near room temperature (RT) are strongly preferred over those appearing only at low temperatures for a variety of technological reasons. The CDW logic gates can have operation principles similar to the spin wave logic gates [10-12] while offering RT operation without utilization of the magnetic field.

In this Letter we report on "graphene-like" mechanical exfoliation of titanium diselenide (TiSe$_2$) thin films and an investigation of their phonon and CDW properties using Raman spectroscopy. Understanding how CDW properties such as the transition temperature change as a function of film thickness is important to any thin-film application. To the best of our knowledge this is the first study of the evolution of Raman spectrum and transition temperature with the thickness of CDW thin films.

TiSe$_2$ belongs to a group of the *van der Waals* materials characterized by the layered crystalline structures. The presence of the van der Waals gaps with weak bonding allows one to exfoliate films with various thicknesses from the corresponding bulk crystals. We have previously exfoliated and studied other van der Waals materials, including individual quintuples of bismuth telluride (Bi$_2$Te$_3$) [13] and atomic tri-layers of titanium ditelluride (TiTe$_2$) [6]. Few-quintuple Bi$_2$Te$_3$ films reveal unusual thermoelectric and topological insulator properties [14-15]. Titanium dichalcogenides TiX$_2$ (X=Se, Te, S) have the hexagonal crystal structure of the space group D$_{3d}$. The TiX$_2$ tri-layers are separated by the van der Waals gaps while inside each layer a Ti atom is surrounded by six chalcogen atoms in the octahedral configuration (Figure 2). These types of crystals with atomic tri-layers as their structural units differ from both graphene – a single atomic layer – and Bi$_2$Te$_3$, which is built of five-fold atomic planes – quintuples. Many transition-metal dichalcogenides reveal transitions to a CDW phase at different temperatures [16-19]. It is conventionally accepted that bulk 1*T*-TiSe$_2$ exhibits the transition to the commensurate (2×2×2) CDW state accompanied by a periodic lattice distortion below *T*≈200 K [20]. The electronic structure of TiSe$_2$ has been extensively studied. However, no consensus has been reached on whether the material is semiconducting [21] or semimetallic [22].

There have been several different models to explain specifically the evolution of the Raman spectrum of the CDW phase. The most common explanation is the formation of a "superlattice" by the $A_u$-symmetry displacement of Ti and Se atoms in the basal plane – perpendicular to the wave vectors in the L-points [20, 23-24]. The high temperature phase with the space group symmetry $D_{3d3}$ contains one formula unit – 3 atoms – per unit cell. The transition results in the formation of the "superlattice" with the space group $D_{3d4}$, which contains eight formula units per unit cell. There are other approaches to intepret the Raman spectrum of the CDW phase. The amplitude mode of the CDW phase is expected to be a Raman active mode. Correspondingly, an amplitudon was introduced as an excitation in the amplitude of CDW [2]. In another model, the multiple Raman modes observed in the CDW phase of the material were attributed to anharmonicity between the amplitude and phase modes of CDW materials [25]. While the mechanisms are not fully understood, signatures of the CDW phase do appear in the Raman spectrum.

In the present work, TiSe$_2$ films were exfoliated onto Si/SiO$_2$ substrates following the standard graphene-like" approach [26]. The thickness, $H$, of the films ranged from the nanometer to micrometer scales. The atomic force microscopy (AFM) inspection (AIST-NT) revealed that surface roughness of our films was $\delta H$~ 1 nm within the investigated regions. The thickness of an individual Se-Ti-Se atomic tri-layer is $H$=0.271 nm. The lattice parameters for TiSe$_2$ are $a$=0.377 nm and $c$=0.6498 nm. As the main characterization technique we used micro-Raman spectroscopy, which allows one to observe the changes in the phonon spectrum near the Peierls transition temperature $T_C$. We also studied the dependence of the transition temperature on the thickness of CDW thin films. The exfoliated films of van der Waals materials are characterized by stronger quantum confinement of charge carriers and phonons compared to films of similar thickness grown by molecular-beam epitaxy grown on lattice matched substrates [13, 27]. One should mention here that the onset of the quantum confinement regime for acoustic phonons starts at much larger thicknesses than that for electrons [28]. We avoided the use of individual tri-layers of TiSe$_2$ in our Raman study owing to strong laser heating even at very small excitation power levels.

The high-temperature phase of TiSe$_2$ belongs to the space group D$_{3d3}$ and contains 3 atoms per unit cell. There are nine zero-center vibrational modes for TiSe$_2$ [29]:

$$\Gamma = A_{1g} + E_g(2) + 2A_{2u} + 2E_u(2). \qquad (1)$$

They include two Raman active $\Gamma$-point phonon modes of $A_{1g}$ in which two Se atoms per unit cell move relative to one another along the $z$ axis and a doubly generate $E_g$ mode in which the Se atoms move opposite to one another along the $x$ or $y$ directions (Figure 2). Both the high-temperature and low-temperature phase of TiSe$_2$ have inversion symmetry. The four odd parity modes of Eq. (1) are divided in two acoustic and two optical modes. The transition to the CDW regime results in formation of a "superlattice" with the space group D$_{3d4}$ which contains 24 atoms – eight formula units – per unit cell.

Raman spectroscopy (Renishaw InVia) was performed in the backscattering configuration under $\lambda$ = 633 nm laser excitation. An optical microscope (Leica) with a 50× objective was used to collect the scattered light. Bulk single crystal TiSe$_2$ is known to have a low thermal conductivity and melting point. The thermal conductivity of the exfoliated thin films is even lower due to the acoustic phonon scattering from the top and bottom surfaces of the sample. For this reason the selection of the excitation power for Raman spectroscopic measurements is important. The power of 0.27 mW on the surface provided meaningful results without local melting or oxidation of the material. No laser damage was observed for the TiSe$_2$ films at this power level. Laser power above 0.5 mW resulted in local melting of the Ti-Se crystal. The local heating for this type of van der Waal material contrasts with that of graphene which has high thermal conductivity [30-31]. Figure 3 shows the Raman spectra of TiSe$_2$ thin films as the excitation power is changed. The observed peaks at 154 cm$^{-1}$, 196 cm$^{-1}$ and 233 cm$^{-1}$ were attributed to $A_{1g}+E_g$, $A_{1g}$ and $E_g$, respectively. The position of the measured peaks at 154 cm$^{-1}$ and 233 cm$^{-1}$ coincides exactly with the calculations for $A_{1g}+E_g$ at the symmetry point M and $E_g$ at the symmetry point L reported previously [29]. The peak at 196 cm$^{-1}$ is close to the reported experimental values of 195 cm$^{-1}$ [23], 187 cm$^{-1}$ [32] and calculated value of 187 cm$^{-1}$ [29]. These studies were consistent in assigning this peak to $A_{1g}$ symmetry.

In order to investigate the evolution of the Raman spectrum of TiSe$_2$ films below and above the transition temperature, we used the samples placed in a cold-hot cell. We varied the temperature in the range from 110 K to 400 K and recorded Raman spectra both in the cooling and heating cycles. Figure 4 shows representative un-polarized Raman spectra from TiSe$_2$ films as the temperature decreases from 400 K to 110 K (cooling cycle). One can clearly see two main peaks at 233 cm$^{-1}$ – 236 cm$^{-1}$ and 196 cm$^{-1}$ – 204 cm$^{-1}$. These peaks can be associated with the $E_g$ and $A_{1g}$ modes of bulk TiSe$_2$, respectively [29,32]. Other prominent features in the spectrum include shoulders and small peaks at 136 – 142 cm$^{-1}$ and 316 cm$^{-1}$. The peaks at ~136 cm$^{-1}$ are due to the $E_g$ phonons [23-32] with contributions coming possibly from the M, A and Γ symmetry points [29]. The broad band at 460-cm$^{-1}$ is likely a second-order Raman scattering peak. The peaks in bulk TiSe$_2$ above ~300-cm$^{-1}$ were attributed to two-phonon processes [32]. The small peak at 316 cm$^{-1}$ was identified as due to $E_g$ phonons. Its energy in TiSe$_2$ bulk crystals was given as 317-cm$^{-1}$ [23] and 314 cm$^{-1}$ [32].

It has been reported that two pronounced phonon modes appear in the bulk CDW phase of TiSe$_2$ at 74 cm$^{-1}$ ($E_g$) and 116 cm$^{-1}$ ($A_g$) [32]. We were not able to distinguish these peaks in the Raman spectra of TiSe$_2$ films at the small laser power levels limited due to the strong local heating effects in the films. However, we did observe reproducible changes in the unpolarized Raman spectrum near the transition temperature of ~200 K. The shoulder at 136 – 142 cm$^{-1}$ becomes a well-resolved peak as the temperature decreases below ~200 K (dark blue line in Figure 4). The second order feature at ~460 cm$^{-1}$ appears only for T<200 K. The data in the heating cycle as the temperature of the samples increase from 110 K to 400 K shown in Figure 5 is largely consistent with the data accumulated in the cooling cycle.

Figure 6 presents Raman spectra of a thinner TiSe$_2$ film (average $H$=100 nm). The main features are still the A$_{1g}$ peak at ~207 cm$^{-1}$ and the E$_g$ peak at 233 cm$^{-1}$. The feature at 316 cm$^{-1}$ is more pronounced and appears as a distinguished peak near the transition temperature. One can also notice that the temperature at which the spectrum modification is observed is shifted to about ~225 K. The intensity of the low-temperature Raman peaks varied from sample to sample. The emergence of the new Raman lines in TiSe$_2$ is often explained by formation of the $2a_0\times 2a_0\times 2c_0$ "superlattice" below the CDW transition temperature [23]. Recording the temperature at which new features in the Raman spectrum appear can provide information on the CDW transition temperature in the exfoliated TiSe$_2$ films.

Figure 7 shows the dependence of the CDW transition temperature on the thickness of the exfoliated TiSe$_2$ thin films. The transition temperature was determined from the onset of the new peaks in the Raman spectrum indicating the new phase. One can see that the temperature increases from ~200 K in the thick films ($H$~2 µm) to ~240 K in the films with the average thickness below ~100 nm. The average thickness range was used in the nanometer scale region owing to the thickness non-uniformity over the laser spot size of about 1 µm (see the inset). From the applications point of view, a possibility of increasing $T_C$ in thin films of CDW materials presents a major benefit. From the physics point of view, the obtained data can be useful for further theoretical studies of the scaling effects on the CDW phase.

It was recently found from density functional theory calculations that both monolayer and bilayer NbSe$_2$ undergo a CDW transition [33]. Moreover, an order-of-magnitude larger energy reduction by transition to a CDW phase in a monolayer of NbSe$_2$ as compared to that in the bulk indicates a much higher transition temperature in the atomically thin films of NbSe$_2$ [33]. A similar study was recently performed for TaSe$_2$, and it was found that the energy reduction from the CDW distortion was similar for both the bulk and the single layer film [34]. It was noted that this difference between NbSe$_2$ and TaSe$_2$ was consistent with the different pressure derivatives of the CDW transition temperature, slightly positive for TaSe$_2$ [35] and negative for NbSe$_2$ [36]. The pressure derivative of the TiSe$_2$ CDW transition temperature is also negative [37], and this is consistent with our experimental finding of increased transition temperature with reduced thickness.

In conclusion, we investigated the temperature and thickness evolution of the Raman spectrum in mechanically exfoliated films of TiSe$_2$. The transition temperature to the charge density wave phase in this material was determined via the appearance of additional peaks in the Raman spectra. It was found that the transition temperature can increase from its bulk value of ~200 K to ~240 K as the thickness of the exfoliated van der Waals films reduces to below one hundred nanometers. This thickness dependence is consistent with the known pressure dependence of the transition temperature. The obtained results add to the knowledge of the change density wave effects and can potentially lead to the use of the CDW collective states as the new state variables in the information processing.

*Acknowledgements*

This work was funded by the National Science Foundation (NSF) and Semiconductor Research Corporation (SRC) Nanoelectronic Research Initiative (NRI) project 2204.001: Charge-Density-Wave Computational Fabric: New State Variables and Alternative Material Implementation (NSF-1124733) as a part of the Nanoelectronics for 2020 and beyond (NEB-2020) program. P.G. thanks Dr. Craig M. Nolen for assistance with micro-Raman spectroscopy.

**FIGURE CAPTIONS**

**Figure 1:** Illustration of the transition from the normal state to the charge-density-wave state below the Peierls temperature $T_C$. The atomic displacements leading to the formation of a "superlattice" results in opening of an electronic energy band gap and modification of the Raman spectrum of CDW materials.

**Figure 2:** (a) Schematic of the crystal structure of $TiSe_2$ – van der Waals layered material; (b) main types of the crystal lattice vibrations in $TiSe_2$.

**Figure 3:** Raman spectra of $TiSe_2$ flakes for different values of the excitation power. The local heating effects even under the moderate excitation power levels are often strong in the van der Waals thin films owing to their low thermal conductivity.

**Figure 4:** Temperature dependence of the $TiSe_2$ Raman spectrum in the cooling cycle from 300 K to 110 K. The main spectral features are indicated in the legends and in the text.

**Figure 5:** Temperature dependence of the $TiSe_2$ Raman spectrum in the heating cycle from 110 K to 400 K. The main spectral features are indicated in the legends and in the text.

**Figure 6:** Temperature dependence of the $TiSe_2$ Raman spectrum of a thinner film ($H$=100 nm) in the cooling cycle from 300 K to 135 K. Note the appearance of well-resolved peaks at ~207 cm$^{-1}$ and ~316 cm$^{-1}$ after the temperature decreases below $T_C$ for a given film thickness. The evolution of Raman spectrum can be used to investigate the dependence of the transition temperature on the thickness of the films.

**Figure 7:** Transition temperature dependence on the $TiSe_2$ film thickness. Inset shows a representative optical image of an exfoliated flake of $TiSe_2$.

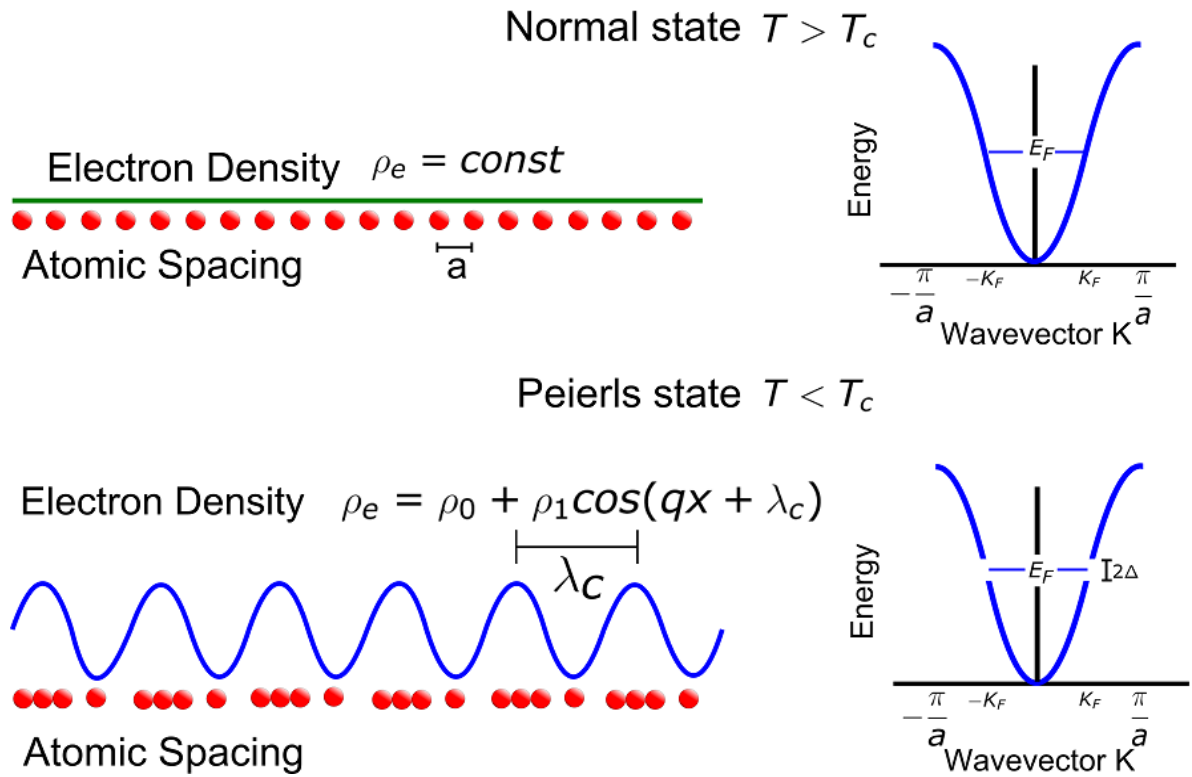

Figure 1

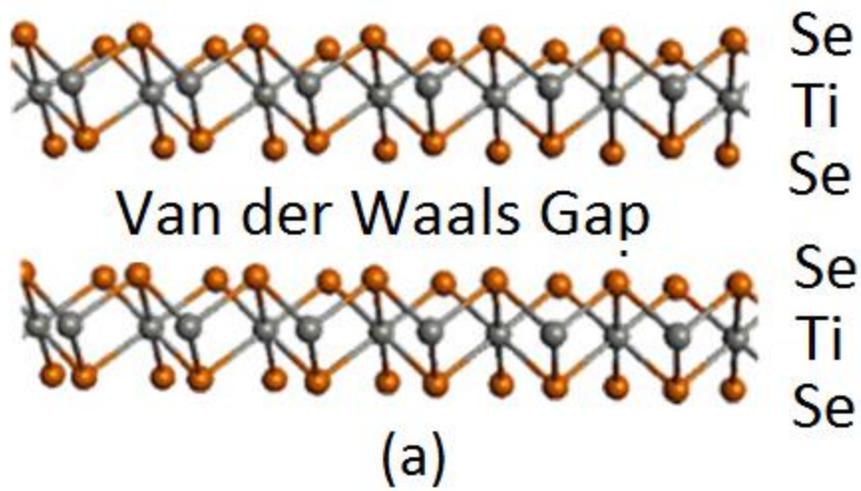

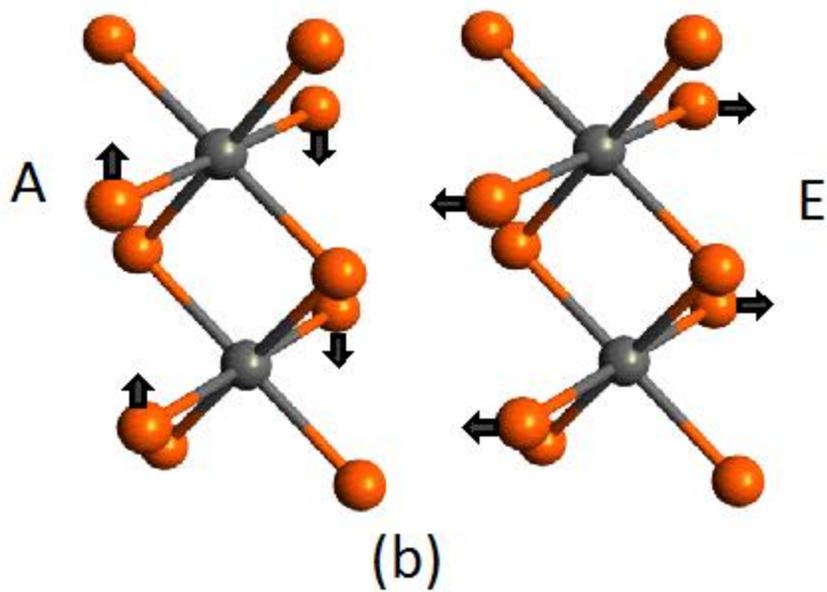

Figure 2

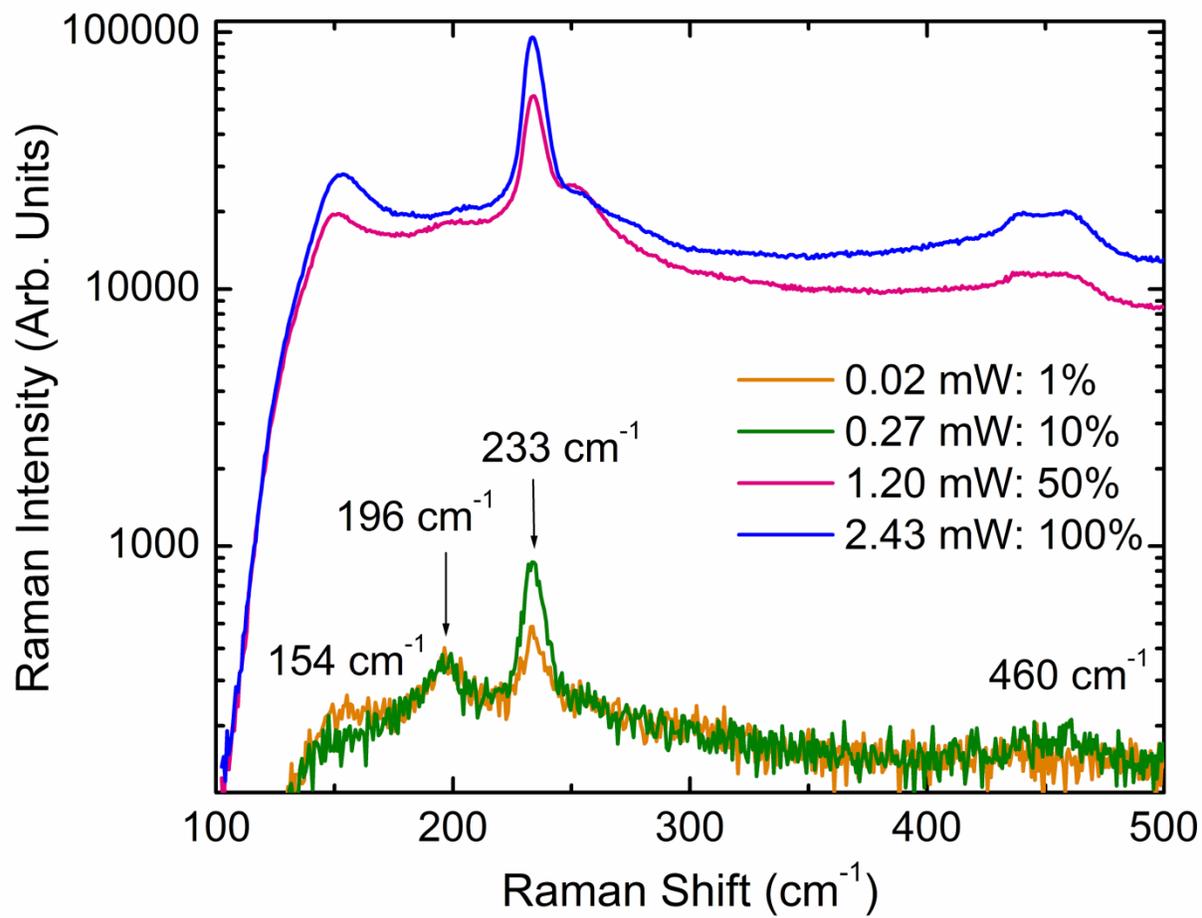

Figure 3

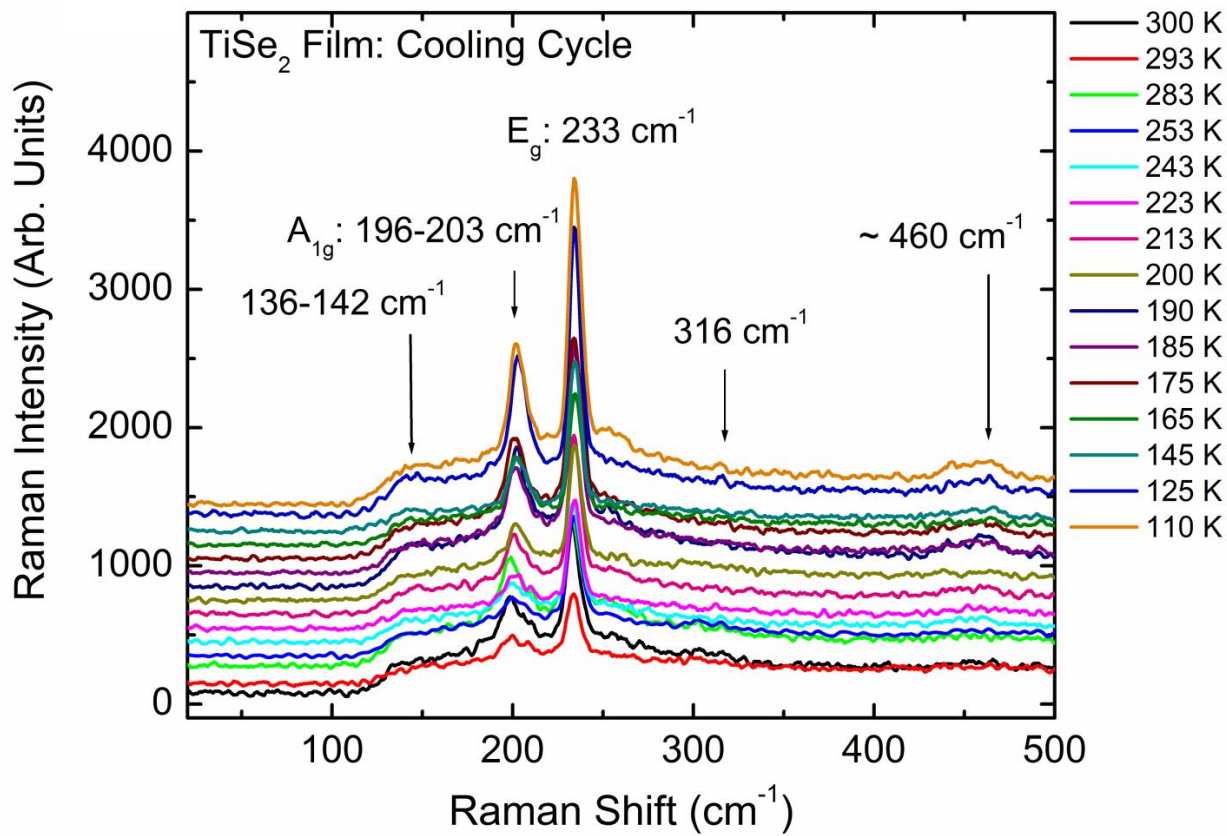

Figure 4

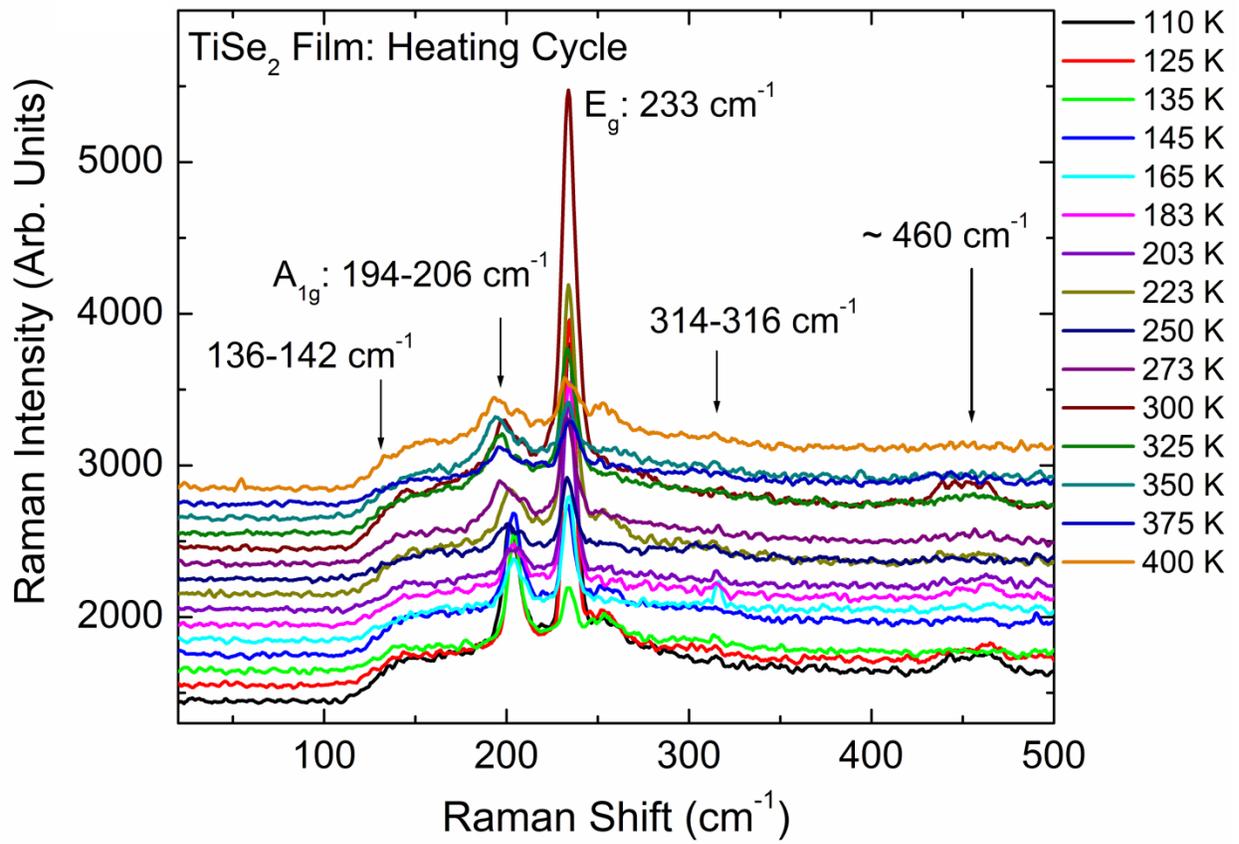

Figure 5

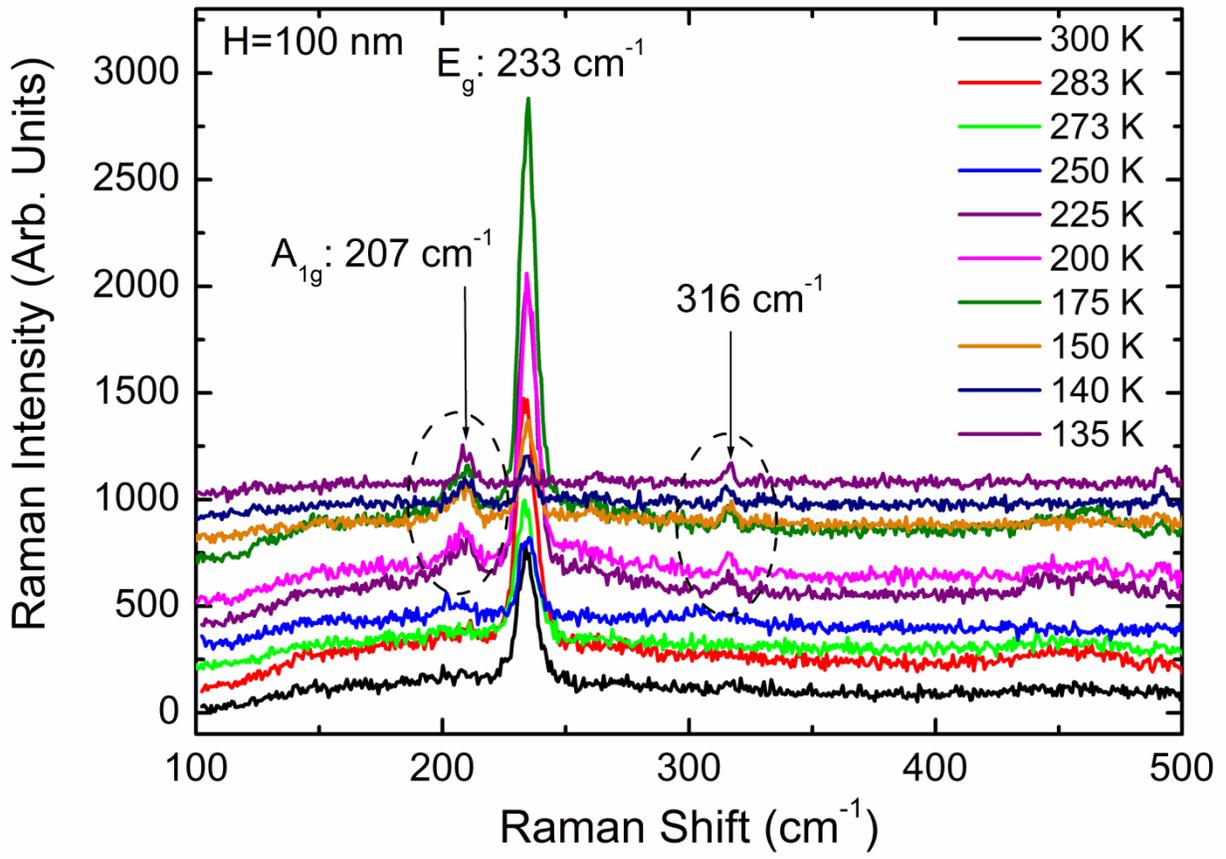

Figure 6

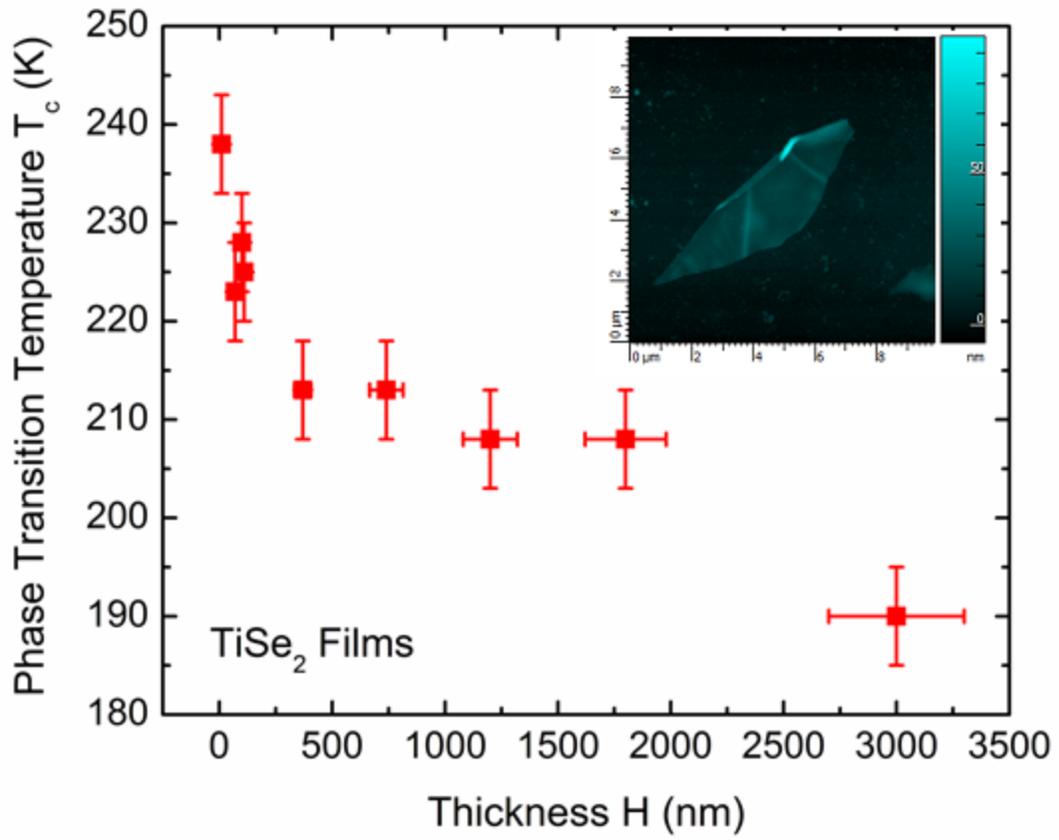

Figure 7